\documentclass[nofootinbib,eqsecnum,showpacs,preprintnumbers,superscriptaddress,floatfix]{revtex4-1}
\usepackage[letterpaper,margin=1in,bmargin=1.5in]{geometry}

\usepackage[utf8]{inputenc}
\usepackage{amsmath,amssymb}
\usepackage{bm} 
\usepackage{CJK}
\usepackage{graphicx}
\usepackage[sort&compress]{natbib}
\usepackage[usenames,dvipsnames,svgnames]{xcolor}

\usepackage{comment}
\usepackage{listings}

\usepackage[pdftex,colorlinks=true,
    linkcolor=blue,
    filecolor=blue,
    urlcolor=blue,
    citecolor=blue,
    plainpages=false,
    bookmarksopen=true]{hyperref} 

\newcommand{\tx}{\lstinline|tx|}

\allowdisplaybreaks

\definecolor{red}{HTML}{A00000}
\definecolor{green}{HTML}{309922}
\definecolor{orange}{HTML}{F68D2E}
\definecolor{purple}{HTML}{7030A0}
\definecolor{mocha}{HTML}{885500}

\makeatletter
\@booleanfalse\titlepage@sw
\makeatother
\begin{document}

\begin{CJK*}{UTF8}{bsmi}

\title{Reducing Disorder: An Information-Theory Formulation of MEV}

\author{Ciaran~\surname{Hughes}}
\email{iamciaran2@gmail.com}

\begin{abstract}
Maximal Extractable Value (MEV) has garnered significant attention in the cryptocurrency community. Such attention is a consequence of the revenue that can be generated from MEV, as well as the risks MEV poses to the fundamental value proposition of the underlying blockchain technology. In this work, we provide an information-theoretic formulation of MEV. With this formulation, we make common statements about MEV mathematically rigorous. For example, we show that i) all non-trivial blockchains and decentralised applications must generate MEV; ii) how MEV can be reduced at the expense of user expressibility; and iii) how MEV can be good or bad from an information theoretic standpoint. 
\end{abstract}

\maketitle
\end{CJK*}

\section{Introduction}
\label{sec:intro}

MEV was originally introduced as Miner Extractable Value in the original {\it{Flash boys 2.0}} \cite{9152675} paper due to the miner's privileged role in building blocks on ETH1. It was hypothesised that the additional revenue from the censorship, reordering, or insertion of transactions during block building could incentivise the centralization of miners. Such a negative externality, it was pointed out, should be avoided due to being diametrically opposed to the decentralized value proposition of the blockchain technology.  

The work of \cite{babel2021clockwork} mathematically quantified single-block miner extractable value as the maximum increase of a miners balance after they build a block. More specifically, using the terminology in that work, take the player $P$ to be a miner, $s$ to be the state of the blockchain prior to the miner building a block, and $b(P,s)$ to be the balance of $P$ in state $s$, then 
\begin{align}
    \text{MEV}(P,s) &= \text{max} \{ b(P,s') - b(P,s) \},
\end{align}
where $s'$ is the state after the mined block. 

However, due to the fact that more players than just miners could extract value from the insertion or re-ordering of transactions, the definition of MEV was updated from miner extractable value to {\it{maximal extractable value}}. Further, this definition includes how other players could impose additional negative externalities on the blockchain ecosystem. Examples of these players include searchers, who look for profitable methods to build transactions (potentially using or exploiting retail users of the blockchain). Depending on the mechanism design of the block building, searchers may be incentivized to spam transactions to the blockchain in order to increase the probability of successfully executing a strategy. This has had the effect of hurting user-experience and driving up gas fees. More recently, the centralization risk from wallets selling users orderflow has been highlighted, leading to a classic payment-for-orderflow (PFOF) model. Moreover, from a moral-hazard perspective, groups of players could be financially incentivized to collude with others, again posing a threat to decentralization. 

A mathematical formalism for multi-domain maximal extractable value, as well as a presentation of the cost of collusion within this formalism, was studied in \cite{obadia2021unity}. For the two blockchain case, the maximal extractable value was mathematically defined to be 
\begin{align}
    \text{MEV}(P,s) &= \max_{\vec{a} \in A_i \cup A_j}\{ev_i(P,s,\vec{a}) + (p_{j\to i})ev_j(P,s,\vec{a})\}  
\end{align}
where $\vec{a} = (a_1,\ldots,a_n)$ is an ordered sequence of valid actions within the set of all valid actions denoted $A_i \cup A_j$, $s$ is the joint state space of the two domains where $s \to s'$ after acting $\vec{a}$ on $s$, $p_{j\to i}$ is a pricing function used to convert the balance of tokens in domain $j$ to a balance on domain $i$, and the extractable value on domain $i$ is defined as 
\begin{align}
ev_i(P, s, \vec{a}) &= b_i(P,s') - b_i(P,s) . 
\end{align}
As illustrated above, the current formulation of MEV is based upon the balances of a players account before and after a sequence of actions. 

More abstractly, Xinyuan Sun {\emph{et.~al.~}}have generalized the formalism of MEV in an mechanism design framework in \cite{3EV}, called 3EV. This framework has generalized MEV into three distinct, yet interacting, categories. First, Monarch MEV, colloquially understood as ``proposer-MEV'', is classified by the extraction of value from the privileged position the proposer/miner has in ultimately deciding the ordering of transactions in a block; second, Mafia MEV, which is colloquially understood to be ``searcher-MEV'' is the extraction of value from non-consensus players using an informational edge to extract value from other players, e.g., arbitragers, sandwich attacks, etc; thirdly, Moloch MEV is the extractable value lost from uncoordination built into the system (conceptually similar to the price of anarchy).

The above definitions of MEV all rely on a players balance increasing after taking a set of actions. However, implicit in each definition was the specific sequence of actions that a player actually took to increase their balance. In the power set of all possible actions, only a small number actually produce a balance increase. In fact, it is purely from an informational advantage that a player can consistently choose the specific sequence of actions that actually increase their balance.  Since MEV/3EV is intrinsically linked to the amount of information available to the different players within a system, it is natural to provide an information-theoretic formulation of MEV. 

Originally developed by Claude Shannon, information theory has proved useful in studying a broad range of fields including communication channels, compression, and encryption (and as such, is a natural extension to study different blockchain behaviour). A description of MEV in terms of information theory can also make colloquial statements about MEV more mathematically quantifiable, and open up a new viewpoint on the phenomena. 

In this direction, it is first important to understand the different players, technologies, mechanisms, and strategies available. We define the different players in Sec.~\ref{sec:players}. Next, in Sec.~\ref{sec:info}, we provide a brief overview of important concepts from information theory. A simple model for atomic arbitrage is also presented in order to illustrate the concepts in Sec.~\ref{sec:model}. In Sec.~\ref{sec:block}, we discuss how different protocols for block building gives rise to different values of MEV. 
The main discussion of our work is in Sec.~\ref{sec:MEV}, where we discuss how this information theoretic formulation of MEV allows us to mathematically quantify common statements about MEV. Lastly,  we finish in Sec.~\ref{sec:conc} with our conclusions and suggestions for further work. 

\section{Who are the Players, And What Are Their Actions?}
\label{sec:players}

Before discussing how information theory is formulated, it is necessary to enumerate some of the different players within the blockchain ecosystem, as well as the actions that are available to them. We briefly present some definitions, and point the reader to \cite{obadia2021unity, babel2021clockwork} for more details.  Initially we will restrict ourselves to a single-domain blockchain that executes transactions sequentially, like Ethereum. However, this framework can be extended straightforwardly to a multi-domain world (which can include a centralized exchange), or a single domain multi-block framework. We also keep the less abstract notations for the ease of understanding for the reader. 

We call the current state of the blockchain domain $s$. A transaction $\tx^{(P)}$ is defined to be any valid action by player $P$ which acts on the current state $s$ according to the state transition function of the blockchain. For later use, we use the notation of a transaction $\tx^{(P)}$ acting on $s$ as $\tx^{(P)} \cdot s = s'$, where $s'$ is the new state after the action has been performed. Notably, an ordered sequence of transactions can be executed sequentially and result in a new state $s' = \tx_i^{(P)}\cdot \tx_j^{(P)} \cdot s$. Further, the allowed actions/transactions of a player depends on the current state of the blockchain. 

In this work we focus on a subset of the allowed players. Each player has a distinct role, and we refer to them as:
\begin{itemize}
    \item Proposer or Builder: The entity responsible for how transactions are ordered within a new block. We refer to a builder in the proposer-builder separation (PBS) context of proof-of-stake (POS) blockchains,  where the builder constructs the block and provides it to the proposer. The proposer then submits a block for consensus. We refer to a proposer in the context of pre-PBS POS blockchains, i.e.,  a proposer both builds and proposers a block. 
    \item Trader or User: An (unsophisticated) entity which performs a single action per transaction, e.g., sends an allowed number of tokens, or swaps tokens in a single pool, both not both within a single transaction. These players can be generalized to perform any number of swaps per transaction, however, for brevity we only take a single swap per transaction in Sec.~\ref{sec:model}. The entities generate arbitrage opportunities, but do not capture them using sophisticated strategies.
    \item Searchers: Sophisticated entities which are attempting to make a profit by using an informational advantage. Searchers commonly perform multiple actions within a single transaction in order to ensure a profit, a scenario known as atomic MEV. Searchers can also make profit using multi-block or cross-domain MEV, which currently do not fall within the purview of atomic MEV\footnote{Once the flashbot's SAUVE chain comes online, it may be possible to have atomic cross-domain MEV.  If blockspace futures become possible then multi-block atomic MEV may be possible.}. 
\end{itemize}

In order to illustrate a minimal information theoretic definition of MEV, these are all players which are necessary. We now move onto reviewing the foundations of information theory. 

\section{What Is Information?}
\label{sec:info}

The most fundamental concept in information theory is the amount of information we can acquire from an event. The informational content of an event $X$\footnote{We assume in this work that the events are discrete, though the formulas can be extended straightforwardly to the continuous case.} is defined to be 
\begin{align}
\label{eqn:info}
I(X) & = -\log(P(X)),
\end{align}
where $P(X)$ is the probability of the event $X$ occurring. As can be seen, the concept of information is intrinsically tied to the probability distribution of $X$. The base of the logarithmic function defines the units of information, and here we take base two, giving units of bits.  

The conceptual understanding of information can be frequently confused. For this reason, we elucidate it further. As an example, take $X$ to be the event of snowflake falling during a snowstorm, and assume $P(X)=1$. In this case, the information we can gain from the event should be zero, since the event tells us no new information. Correctly, $I(X) = 0$ using (\ref{eqn:info}). However, if $X$ is now taken to be a snowflake falling in the desert on a warm day, then the probability of this event is small, $P(X) \approx 0$. Using (\ref{eqn:info}) again, we can see this event has a large amount of information which we can acquire upon observation. For these reasons, information is sometimes called the amount of "surprise" of observing an event in the field of natural language processing, where a low information event with $P(X)=1$ is not surprising, and $P(X)\approx0$ is very surprising. Information is also know as how certainty or uncertain we are about $X$.

Another fundamental concept is known as entropy. Entropy is the expectation of information of a random variable $X$, defined as 
\begin{align}
\label{eqn:entropy}
    H(X) &= \mathbb{E}[I(X)] = - \sum_x p(x) \log(p(x)).  
\end{align}
Entropy should be understood to be the average amount of information in a system, or equivalently, the average amount of surprise we would expect upon observing the system. For the purposes of this work, it is preferable to understand entropy in the physics context: as the average amount of disorder or volatility in a system. As such, $H(X) \ge 0$. Entropy also has units of bits, and can be understood as the number of bits needed to encode the disorder of the system. This understanding will be particularly useful when applying entropy or information to MEV. 

In the terminology of information theory, an action or transaction is an event, which follows some probability distribution. 
We can describe an ordered sequence of $n$ transactions $\vec{\tx} = (\tx_n, \ldots, \tx_1)$ acting on a state $s$ as a stochastic process. The resulting blockchain state $s' = \tx_n \cdots  \tx_1 \cdot s = \vec{\tx} \cdot s$ is the result of the stochastic process, where we have introduced the shorthand $s'=\vec{\tx} \cdot s$. We are interested in the disorder of the state after these $n$ transactions.\footnote{We are not interested in the disorder of the $n$ transactions, just of their effect on the state. } 

In order to define MEV as some function of entropy, we would conceptually like to know the change in disorder of the blockchain state as sequential transactions are included. Intuitively, it is useful to  study the conditional entropy after including more transactions acting on a general state $s$. The conditional entropy can be written as 
\begin{align}
    \label{eqn:changeentropy}
     H(\tx_2 \cdot \tx_1 \cdot s| \tx_1 \cdot s) &= H(\tx_2 \cdot \tx_1 \cdot s) - H(\tx_1 \cdot s). 
\end{align}
$H(\tx_2 \cdot \tx_1 \cdot s| \tx_1 \cdot s)$ tells us whether or not the inclusion of the single new transaction $\tx_k$ will increase the disorder of the blockchain state $s$. 

Therefore, by repeatedly applying (\ref{eqn:changeentropy}), the average entropy of the state after $n$ transactions can be written as 
\begin{align}
    \label{eqn:jointentropy}
    H(\vec{\tx}\cdot s) &= \sum^n_{k=2}  H(\tx_k \cdots  \tx_1 \cdot s | \tx_{k-1} \cdots  \tx_1 \cdot s)  + H(\tx_1 \cdot s),
\end{align}
where $H(X|Y)$ is the conditional entropy of $X$ given $Y$. Eqn.~(\ref{eqn:jointentropy}) makes intuitive sense since it indicates that the disorder in the state after $n$ transactions is just the initial disorder and the sum of the changes of disorders after each transaction is applied. We will make use of this later. 
Eqn.~(\ref{eqn:jointentropy}) tells us the total disorder (or volatility) in the blockchain state after applying the $n$ transactions. Taking $n$ to be the size of the block, this can be useful in quantifying how blocks are built, and from which we can derive mathematical statements about generic blockchains, as discussed in Sec.~\ref{sec:block}. 

With this framework provided, we can move onto illustrating these concepts in a model. After some intuition is built, we can make general statements about MEV in information-theoretic terms. 

\section{An Information Theoretic Model For MEV}
\label{sec:model}

While the specific choice of model does not change the conclusions of this work, for illustrative purposes, we study a simple blockchain model.  Assume that the blockchain state consists solely of account balances and two AMM pools which can be arbitraged. For brevity, we denote the state as $s=[p_1, p_2]$, where $p_i$ is the $i^{\text{th}}$ pool and  contains the blockchains native token. We take the pools to initially be in equilibrium, and as such the initial condition of the state entropy (or amount of disorder) to be $H(s=[p_1,p_1])=0$. We define two pools in equilibrium as having the same number of tokens, or indistinguishable. 

Next, we need to consider the event spaces of the different players. As discussed in Sec.~\ref{sec:players}, we will study scenarios where traders only perform a single swap within a transaction. Without loss of generality, we can assume that these traders will only interact with the first pool\footnote{In this model, only the difference between the pools matters, and so we can assume only the first pool is traded on. Equivalently, we can define $A_i$ to be one of the tokens in the pool, then the state difference of the pools can be defined as $S = N^{(1)}_{A_1} - N^{(2)}_{A_1}$, where $N^{(j)}_{A_1}$ is the number of $A_1$ tokens in pool $j$. }.

In this model, the traders can either buy one single unit of the native token with probability $p$, or sell one single unit of the native token $q=1-p$, according to a standard Bernoulli distribution.\footnote{A Bernoulli random variable is normally presented as a random variable $X$ which can have values $1$ or $0$ probabilities $p$ and $q$ respectively. We have shifted the $0$ value to $-1$ for our purposes. } After a single trade, the state will be out of equilibrium by a single unit.  After $2$ consecutive trades, the state space can either be two units out, zero units out, or minus two units out, with probabilities given by $p^2, 2pq,$ and $q^2$ respectively. In general, after $n$ consecutive trades, the state space is the random variable given by $\Delta s_n = \sum^n_{i=1} \tx_i$, where each $\tx$ can be thought of as the Bernoulli random variable described above. $\Delta s_n$ describes how many units out of equilibrium the state space is after the $n$ transactions, and is distributed according to a Binomial distribution (as it is the sum of $n$ Bernoulli random variables). The possible values for $\Delta s_n$, which defines its event space, are $n-2a$, where $a=0,\ldots, n$ are the number of sell orders. As such, the probability of observing $\Delta s_n$ equal to some value $k=n-2a$ is given by $P(\Delta s_n=n-2a) = {n\choose a}p^{n-a}q^a$.

Imagine a scenario where $p=0.5$, i.e., its equally likely for traders to buy or sell. Intuitively, based on the dynamics of the traders in this model, after $n$ trades there is a large variation in the possible outcome of the state, and hence a large uncertainty in what the state could be. $p=0.5$ maximizes the disorder or entropy, as there is no preference for selling or buying in this model. Equivalently, there is maximum uncertainty on how the traders actually trade, and as such, a maximum uncertainty on what the state will be. 

However, if $p$ is close to one, so that traders are buying more than selling, then after $n$ transactions, we would expect the dynamics to ensure there is a higher probability of the pools to be out of equilibrium. There is still entropy in this case, yet, it is less than the $p=0.5$ scenario. In this case, there is less disorder or less uncertainty in what the traders are doing, and as a consequence, less uncertainty on the outcome of state of the blockchain. For reference, the formula for the entropy of a binomial distribution is $0.5\log_2[2\pi e np(1-p)] + \mathcal{O}(1/n)$. 

Next, if the two pools are trading at different prices and there exists disorder in the system, then a searcher can realize a profit by arbitraging them. Some subtleties of the definition of profit are needed here. In the current work, we define profit to be atomically, that is an increase of searchers balances within a single transaction (where a bundle is considered a single transaction). It is possible to extend this definition to include multi transactions across different blocks, and we touch on this topic in our conclusions. 

Whereas the traders actions were independent from the state of the blockchain, the searchers will not be independent. A reasonable model of the searchers arbitrage strategies are that they will always take this profit if a pool is out of equilibrium, i.e.,  $P(\tx^{(S)} | s)$ is equal to $0$ if $s$ is in equilibrium, and equal to $1$ otherwise. We assume there are no gas or swap fees, so that a pool can be completely equilibrated. As such, the change in entropy from a searchers transaction being executed on a state $s$ which is out of equilibrium is given by Eqn.~(\ref{eqn:changeentropy}) and is  $H(\tx^{(S)} \cdot s') = 0$.\footnote{If an arbitrage is performed on an equilibrium state, then we define $\tx^{(S)}\cdot s = s$,  so that the change in entropy is zero. This is equivalent to if the transaction did not occur at all. } This is an important observation: whereas including a traders transaction can only increase the disorder of the state, a searchers arbitrage transaction will always decrease the disorder. This is an important distinction between these different players. To generalize this statement further, we make the following definition. \\
\newline
{\bf{Definition: }} Atomic MEV is any atomic transaction which decreases the entropy (or disorder) of the blockchain state. \\
\newline
{\bf{Corollary: }} Decreasing the entropy of the blockchain state, which is atomic MEV, costs resources, and so will only be performed if there is an economic incentive to do so. As such, atomic MEV leads to an increase in the players balance. \\

Above, we have defined atomic MEV purely in terms of information theory concepts. Moreover, we have related how the information theory definition naturally leads to the definitions of MEV in terms of the balances of players. This connects our definition to those that are currently in the literature. 

To distinguish between other classes of MEV which occur, we provide one further definition: \\
\newline
{\bf{Definition: }} Bad atomic MEV is any atomic transaction which increases the balance of a player but increases the entropy (or disorder) of the blockchain state. \\

Implicitly, we have defined atomic MEV to be {\emph{good}} atomic MEV. A successful arbitrage fits into the domain of good MEV, since it reduces the disorder of the blockchain state. However, an exploit of the pools, which increases the balance of the player but would increase the disorder on the state, would not count as good MEV. 
Effectively, a searcher performing good MEV is performing some benefit to the blockchain by decreasing the disorder of the state, and as such, is compensated financially for their service.

\section{Different Block Building Mechanism Produce Different Entropy}
\label{sec:block}

The above definition of good MEV relies on the concept of decreasing disorder or entropy. Since the amount of disorder in the blockchain depends on both the number and ordering of users transactions, the amount of MEV is also a function of both. The number and ordering of transactions is determined by the block building procedure. We illustrate different block building mechanisms for the model described in Sec.~\ref{sec:model}, and showcase the different amounts of extractable value in each case. This illustrates how the blocks are built effects the entropy of a blockchain.

We take $\tx_i$ to be i.i.d random variables representing traders, as described in the previous section. We denote the searchers transactions as $\tx^{(S)}$. We also assume that there are a total of $M$ transactions in the mempool, which an entity can use to build a block of size $n$. 

\subsection{Min Oracle MEV}

We define Min Oracle MEV to be the scenario where the proposer can choose the exact $n$ transactions which minimizes the good MEV available when building a block. With a brute force approach, there are ${M \choose n}$ different blocks possible. As a general block building protocol, such an approach is unrealistic due to the high complexity involved, hence why an Oracle is needed. 

However, we can examine the typical payoff from good MEV in this scenario. This oracle would match subsets of buy trades which would organically and randomly execute against equal subsets of sell trades. This is conceptually the idea behind matching transactions based on the ``coincidence-of-wants'' implemented in the CowSwap protocol\footnote{\href{https://swap.cow.fi/}{https://swap.cow.fi/}}.  If the mempool is sufficiently large, then the block would consist of alternating buy and sell orders after each other. 

Thus, after oracle proposer blocks are built, the state is distributed according to 
\begin{align}
\label{eqn:orcale}
\Delta s  = \min_{\{\tx_j\}} \sum^{\lvert \tx_j\rvert}_{k=1} \tx_k  ,  
\end{align}
where we are minimizing over any subset of transactions which minimizes every partition of the sum. 

In this scenario, the expected payoff for good MEV is zero (if $n$ is even), and since the searcher is not providing any utility by decreasing the disorder of the state, this is to be expected. If any block builder did not adhere to minimising the entropy, it would highlight the moral hazard problem: such an entity would be incentivized to create a more disordered blockchain state, and then insert their own transaction to profit from reducing the disorder. However, this would not be incentive compatible with the ethos of an impartial decentralized network. As such, the role of reducing disorder should be outsourced to another player (as is currently done during in a protocol-builder separation framework). 

\subsection{FCFS Blocks}

In a first-come-first-serve (FCFS) block building scenario, the pay-off for MEV is different still. Discussing the fairness of FCFS protocols and their issues related to latency wars are outside the scope of this work, yet we assume that a reasonable strategy for searchers would be to constantly spam arbitrages in the mempool. FCFS would decrease the overall disorder of the pool, as the market dynamics would enforce the block builder to alternate user transactions with searchers arbitrages, and the resulting disorder would be kept to a minimum by the market. In this case, the expected payoff for MEV is $\lfloor n /2 \rfloor$ times the payoff for a single arbitrage.

\subsection{Closed-Bid Dutch Auction}

In the MEV-boost implementation of block building, players would submit either a single valid transaction or bundles of valid transactions to include in a block, and every bundle would provide a tip. The builder ranks the tips, and includes the top $n$ transactions/bundles that tip the most (and which do not revert). If we assume the size of the tip is proportional to the MEV transaction profits, then in a competitive searcher market the larger MEV transactions will be associated with larger tips, then the blocks would be built with bundles which decrease the disorder of the state in descending order. 

In summary, it is not practically possible for builders to combinatorially compute blocks which maximize or minimize MEV, and nor is it necessarily a good framework due to the temptation to censor or include transactions for personal profit. As such, it is better to outsource the entropy reduction of the blockchain state to external players, which are called searchers, and for those players to form a competitive marketplace and get paid to reduce this entropy of the blockchain state. As it is ultimately users which generate this disorder, there is a reason to believe the searchers should share some profits with the users (without which, there would be no profit possible).

\section{All Blockchains Induce MEV}
\label{sec:MEV}

In Section \ref{sec:info} we formulated an information theoretic description of a sequence of transactions. In Section \ref{sec:model} we gave an intuitive picture of this description using a model, and there gave a definition of good MEV in information theoretic terms. The exact model of the distributions were not important, and we can now make a general claim about MEV. With the definition of good MEV, and the general definition of entropy, \\

{\bf{Corollary: }} No non-trivial blockchain can be constructed without MEV. \\ 

Adding increasingly more stochastic information to the blockchain would only further increase the disorder. No blockchain can avoid this property while also having utility\footnote{Unless the utility of the blockchain was to only allow a pre-determined transaction type, or to be a stochastic random event generator.}. Without MEV, the entropy (disorder) of the blockchain would always increase due to the second law of thermodynamics. 

With good MEV, this disorder is reduced, allowing blockchains to be generically useful. As illustrated in Sec.~\ref{sec:model}, entropy is maximized when we are more uncertain about what a users transaction will be. Broadly speaking, the more expressible a user can be on a blockchain, the more disorder/entropy there will be. And as a proxy, the more good MEV there will be as a consequence. 

Reducing the MEV surface amounts to reducing the overall entropy of a blockchain, which amounts to reducing the expressibility of users transactions. From a utilization standpoint, reducing users ability to express themselves has diminishing returns. As such, it stands to reason that it is not very fruitful to overly optimize the reduction of MEV, but rather to optimize how the creators of the MEV (the users themselves) can realize some of the profits which they are creating. In this world, the users are still able to express themselves fully, while also getting some rewards from the MEV they generate. 

\section{Conclusions}
\label{sec:conc}

In this work we have formulated transactions within a blockchain as a stochastic process. Doing so has allowed us to formulate MEV in the language of information theory in Sec.~\ref{sec:info}. In Sec.~\ref{sec:model}, we illustrated these concepts by studying the entropy or disorder within a blockchain state. 

The first key takeaway from this work is our information-theoretic definition of (good) atomic MEV, which is {\emph{any atomic transaction which decreases the entropy of the blockchain state}}. Notably, we do not make reference to players balances in our definition of MEV. In Sec.~\ref{sec:block}, we examined the expected MEV payoff under various block builder schemes. Here we have shown that a proposer could maximize MEV or minimize MEV depending on their prescription of building. 

The second key takeaway from this work is showing that every non-trivial blockchain must induce good MEV. Notably, since users transactions introduce entropy into the system, without good MEV, entropy within the blockchain must grow. Such a blockchain would have no utility (outside of being a random state machine). Good MEV is needed to reduce the entropy of the blockchain state. Reducing disorder could be a function which the builder performs, tying into the concept of Monarch MEV. However, due to the moral hazard problem of allowing builders to potentially censor transactions, the reduction of entropy is exported to external entities called searchers. Consequently, searchers are compensated for their service from MEV profits. 

Since MEV is related to the amount of disorder that exists within a blockchain, and that this disorder is a stochastic process arising from users transactions, reducing the MEV surface equates to reducing the ability of users to express themselves. However, reducing users ability to express themselves is not a viable solution for a highly utilizable blockchain. As such, a likely middle ground is allowing users to express themselves appropriately (hence creating a good MEV environment), but returning some of the profits to them from the disorder that they are generating. 

The breadth of future work in this direction is wide reaching. In terms of privacy, an environment in which there are private transactions will increase the entropy of the state, and hence give rise to more good MEV. In terms of cross-domain MEV (including where one domain can be a centralized exchange), collusion between two sequencers can reduce the disorder of the state, and hence reduce the amount of good MEV between the domains. In this regard, shared information can be tied further in the concept of Moloch MEV, and the MEV arising from uncoordination. In terms of multi-block MEV, the definition of atomic good MEV would need to be extended to include a non-atomic scenario. All these scenarios could yield interesting insights into the nature of MEV, and help reduce the negative externalities associated with MEV. 

\bibliography{paper.bib}

\begin{thebibliography}{4}%
\makeatletter
\providecommand \@ifxundefined [1]{%
 \@ifx{#1\undefined}
}%
\providecommand \@ifnum [1]{%
 \ifnum #1\expandafter \@firstoftwo
 \else \expandafter \@secondoftwo
 \fi
}%
\providecommand \@ifx [1]{%
 \ifx #1\expandafter \@firstoftwo
 \else \expandafter \@secondoftwo
 \fi
}%
\providecommand \natexlab [1]{#1}%
\providecommand \enquote  [1]{``#1''}%
\providecommand \bibnamefont  [1]{#1}%
\providecommand \bibfnamefont [1]{#1}%
\providecommand \citenamefont [1]{#1}%
\providecommand \href@noop [0]{\@secondoftwo}%
\providecommand \href [0]{\begingroup \@sanitize@url \@href}%
\providecommand \@href[1]{\@@startlink{#1}\@@href}%
\providecommand \@@href[1]{\endgroup#1\@@endlink}%
\providecommand \@sanitize@url [0]{\catcode `\\12\catcode `\$12\catcode
  `\&12\catcode `\#12\catcode `\^12\catcode `\_12\catcode `\%12\relax}%
\providecommand \@@startlink[1]{}%
\providecommand \@@endlink[0]{}%
\providecommand \url  [0]{\begingroup\@sanitize@url \@url }%
\providecommand \@url [1]{\endgroup\@href {#1}{\urlprefix }}%
\providecommand \urlprefix  [0]{URL }%
\providecommand \Eprint [0]{\href }%
\providecommand \doibase [0]{http://dx.doi.org/}%
\providecommand \selectlanguage [0]{\@gobble}%
\providecommand \bibinfo  [0]{\@secondoftwo}%
\providecommand \bibfield  [0]{\@secondoftwo}%
\providecommand \translation [1]{[#1]}%
\providecommand \BibitemOpen [0]{}%
\providecommand \bibitemStop [0]{}%
\providecommand \bibitemNoStop [0]{.\EOS\space}%
\providecommand \EOS [0]{\spacefactor3000\relax}%
\providecommand \BibitemShut  [1]{\csname bibitem#1\endcsname}%
\let\auto@bib@innerbib\@empty
\bibitem [{\citenamefont {Daian}\ \emph {et~al.}(2020)\citenamefont {Daian},
  \citenamefont {Goldfeder}, \citenamefont {Kell}, \citenamefont {Li},
  \citenamefont {Zhao}, \citenamefont {Bentov}, \citenamefont {Breidenbach},\
  and\ \citenamefont {Juels}}]{9152675}%
  \BibitemOpen
  \bibfield  {author} {\bibinfo {author} {\bibfnamefont {P.}~\bibnamefont
  {Daian}}, \bibinfo {author} {\bibfnamefont {S.}~\bibnamefont {Goldfeder}},
  \bibinfo {author} {\bibfnamefont {T.}~\bibnamefont {Kell}}, \bibinfo {author}
  {\bibfnamefont {Y.}~\bibnamefont {Li}}, \bibinfo {author} {\bibfnamefont
  {X.}~\bibnamefont {Zhao}}, \bibinfo {author} {\bibfnamefont {I.}~\bibnamefont
  {Bentov}}, \bibinfo {author} {\bibfnamefont {L.}~\bibnamefont {Breidenbach}},
  \ and\ \bibinfo {author} {\bibfnamefont {A.}~\bibnamefont {Juels}},\ }in\
  \href {\doibase 10.1109/SP40000.2020.00040} {\emph {\bibinfo {booktitle}
  {2020 IEEE Symposium on Security and Privacy (SP)}}}\ (\bibinfo {year}
  {2020})\ pp.\ \bibinfo {pages} {910--927}\BibitemShut {NoStop}%
\bibitem [{\citenamefont {Babel}\ \emph {et~al.}(2021)\citenamefont {Babel},
  \citenamefont {Daian}, \citenamefont {Kelkar},\ and\ \citenamefont
  {Juels}}]{babel2021clockwork}%
  \BibitemOpen
  \bibfield  {author} {\bibinfo {author} {\bibfnamefont {K.}~\bibnamefont
  {Babel}}, \bibinfo {author} {\bibfnamefont {P.}~\bibnamefont {Daian}},
  \bibinfo {author} {\bibfnamefont {M.}~\bibnamefont {Kelkar}}, \ and\ \bibinfo
  {author} {\bibfnamefont {A.}~\bibnamefont {Juels}},\ }\href@noop {} {\enquote
  {\bibinfo {title} {Clockwork finance: Automated analysis of economic security
  in smart contracts},}\ } (\bibinfo {year} {2021}),\ \Eprint
  {http://arxiv.org/abs/2109.04347} {arXiv:2109.04347 [cs.CR]} \BibitemShut
  {NoStop}%
\bibitem [{\citenamefont {Obadia}\ \emph {et~al.}(2021)\citenamefont {Obadia},
  \citenamefont {Salles}, \citenamefont {Sankar}, \citenamefont {Chitra},
  \citenamefont {Chellani},\ and\ \citenamefont {Daian}}]{obadia2021unity}%
  \BibitemOpen
  \bibfield  {author} {\bibinfo {author} {\bibfnamefont {A.}~\bibnamefont
  {Obadia}}, \bibinfo {author} {\bibfnamefont {A.}~\bibnamefont {Salles}},
  \bibinfo {author} {\bibfnamefont {L.}~\bibnamefont {Sankar}}, \bibinfo
  {author} {\bibfnamefont {T.}~\bibnamefont {Chitra}}, \bibinfo {author}
  {\bibfnamefont {V.}~\bibnamefont {Chellani}}, \ and\ \bibinfo {author}
  {\bibfnamefont {P.}~\bibnamefont {Daian}},\ }\href@noop {} {\enquote
  {\bibinfo {title} {Unity is strength: A formalization of cross-domain maximal
  extractable value},}\ } (\bibinfo {year} {2021}),\ \Eprint
  {http://arxiv.org/abs/2112.01472} {arXiv:2112.01472 [cs.CR]} \BibitemShut
  {NoStop}%
\bibitem [{3EV()}]{3EV}%
  \BibitemOpen
  \href@noop {} {\enquote {\bibinfo {title}
  {https://hackmd.io/3poiwjbord-mjozm4oujww},}\ }\bibinfo {howpublished}
  {\url{https://hackmd.io/3pOIwjbORd-MJOZM4OUJWw}},\ \bibinfo {note} {accessed:
  2023}\BibitemShut {NoStop}%
\end{thebibliography}%

\end{document}